# CROWDSOURCED DATABASES

# AND SUI GENERIS RIGHTS

Gonçalo Simões de Almeida[*]

Gonçalo Faria Abreu[**]

## Abstract

In this study we propose a new concept of databases (crowdsourced databases), adding a new conceptual approach to the debate on legal protection of databases in Europe. We also summarise the current legal framework and current indexing and web scraping practices - it would not be prudent to suggest a new theory without contextualising it in the legal and practical context in which it is developed.

**Keywords:** databases; crowdsourced databases; data; *sui generis* rights; indexing; scraping; information.

**Index:** 1. Introduction; 2. Database Protection; 3. Crowdsourced Databases.

[*] Lawyer, Master in Law and Management (e-mail: simoesdealmeida.g@gmail.com).

[**] Engineer specialised in machine learning, Master in Electrotechnical and Computer Engineering (e-mail: goncalo.faria.abreu@tecnico.ulisboa.pt).

# 1. Introduction

## 1.1. Online Information Society

More than an information society, we currently live in an online and interconnected information society: information, spread through various sources, is largely accessible on websites. Manual search would be extremely inefficient, to the point that an entire human life would be insufficient to analyse all the relevant information, let alone, to sort it out or organise it.

In fact, human society has always structured, recorded and stored information[1], starting with language, then evolving into writing and taking on the dimension we now know (or think we know). With the birth of the internet, physical barriers were broken down and the cost of storing information today tends to be zero. Similarly, the requirements for its access and registration are now completely different. Thus, it is no surprise that, in 2019 alone, the so-called Data Economy[2], has surpassed the estimated amount of 400 billion euros for the 27 Member States of the European Union together with the United Kingdom, still showing a clearly upward trend[3].

Likewise, access to information and the use of databases that aggregate it, has evolved. Techniques such as indexing and web scraping are widely used today. One cannot move on to the legal framework without knowing them, therefore, we shall devote the next paragraphs to these technological tools.

Indexing is the tool through which it is possible to distinguish relevant information from the globality of existing information, for instance, in a certain book or platform. In the book example, as a set of information distributed over a collection of pages, indexing is done by the index. An index may be something simple, following an alphabetical order (or in the case of a book: the pages), or something complex such as using contexts of the researched object.

---

[1] Since this is not the object of our study, we refer to language (sign language and verbal) as the first of these systems. In fact, as long as humanity exists, there has been information processing and storage.

[2] For the purposes of this text, the Data Economy corresponds to the general impacts of the Data Market on the economy as a whole. Thus, this concept is used in the study on which we based ourselves (see the following footnote).

[3] As per Cattaneo, Gabriella, *et al.*; *The European Data Market Monitoring Tool - Key Facts & Figures, First Policy Conclusions, Data Landscape and Quantified Stories*; Publications Office of the European Union; Luxemburg, 2020. A study commissioned by the European Commission that is based on evidence and facts on the European Union's data economy, with the aim of measuring policy progress in this economic area within the overall framework of the Digital Single Market Strategy. Other conclusions: i) the value of the Data Market alone is estimated at 75 billion euros; ii) the Data Economy has an impact of 2.6% on the Gross Domestic Product of the Member States of the European Union.

With the exponential growth of digitally stored information, accessible by everyone online, the use of a simple index appears unfeasible. This is how we move from simple ordering to understanding contexts: using a set of statistical algorithms, often called machine learning, it becomes possible to analyse a set of data to conduct "intelligent" research through a similarity metric. For example, when searching a set of images using the word "banana", an index capable of analysing context, perceives the word in its genesis and uses computer vision analysis techniques to find objects in images similar to a banana[4]. Taking documents as an example, we replace page indexes with search engines that search for documents based on the context of what was written[5]. Statistical learning was the step taken in an attempt to bring, through the use of statistical standards, the information search tool (in this case, a search engine) closer to the process of filtering data by relevance, as done by human intelligence. The basis of the context we use is the data that portrays our researched object, it is with this data that we create indexes with context.

These techniques represent types of use that were unthinkable twenty years ago, such is the demand that the current dimension of available information places on us. Humanity's digitally stored information grows approximately an order of magnitude every 10 years[6].

If indexing is a way to organise information, web scraping (data scraping; scraping) is the tool that allows obtaining data on a large scale. Simply put, web scraping is carried out through an automated visit to publicly accessible websites[7] on the internet, unstructured, without resorting to indexing and in an unsupervised manner, meaning, without human intervention. This is made possible through simple computer programs, that visit publicly available websites and store their content for further analysis.

---

[4] A Zhou, *et al.*; *Recent Advance in Content-based Image Retrieval: A Literature Survey*; 2017, available at https://arxiv.org/pdf/1706.06064.pdf.

[5] Devlin, Jacob, *et al.*; "Bert: Pre-training of deep bidirectional transformers for language understanding" *in The 2019 Conference of the North American Chapter of the Association for Computational Linguistics: Human Language Technologies Proceedings of the Conference - Vol. 1 (Long and Short Papers)*; 2019, available at https://www.aclweb.org/anthology/N19-1423.pdf.

[6] Hilbert, Martin and López, Priscila; *The world's technological capacity to store, communicate, and compute information*; 2011; pages 60 to 65, available at https://science.sciencemag.org/content/332/6025/60.

[7] By "publicly accessible" we mean websites listed on the Domain Name System (DNS) without being protected by a password or login that requires a user's identification. The specification of this distributed system is set out in "Domain Names - Implementation and Specification";

Scraping may also be used in the context of indexing, to obtain online data - as a tool for extracting information - in order to allow statistical and machine learning algorithms to effectively *learn* their context.

Finally, it is important to recognise that the scraping tool may be used in two ways: "symbiotic", corresponding to the use to build innovative indexes that benefit both parties; and abusive or "parasitic" [8], when it takes advantage of the data source, to its detriment. Beyond the standards, we can confidently say that only agents who use scraping symbiotically survive the test of time.[9]

## 1.2. Websites as Protected Databases

As we saw in the previous point, indexing, as a technique for structuring and organising databases, and scraping, as a data collection tool, allow the user to access, in a timely manner[10], the entire internet, collecting and sorting those websites that the user "visits" on a database. As an example, we can think of search engines or surveys that require the analysis of multiple websites[11].

---

available at https://www.ietf.org/rfc/rfc1035.txt. DNS terminology is set out in "DNS Terminology"; available at https://tools.ietf.org/html/rfc8499.

[8] The negative connotation of the term "parasitism", as malicious conduct (by reference to biology), induces a prior reproach. We prefer - albeit contrary to the generality of the doctrine - the term "abuse". So we will.

[9] Hirschey, Jeffrey; "Symbiotic relationships: Pragmatic Acceptance of Data Scraping"; *in Berkeley Technology Law Journal, Vol. 29, 2014*; 2014; pages 897 *et seq.*

[10] With the evolution of internet capacity, good collection time tends to be zero. As such, good time shall be interpreted as a rate of visits that respects the computational resources of the visited website.

[11] The European Commission has already used this technique to assess compliance with standards for consumer protection on websites in the test it carried out on traders to assess compliance with Directive 2013/11/EU of the European Parliament and of the Council, of May 21st, 2013, on alternative resolution of consumer disputes, amending Regulation (EC) No. 2006/2004 and Directive 2009/22/EC and Regulation (EU) No. 524/2013 of the European Parliament and of the Council, of May 21st, 2013, on online consumer dispute resolution, amending Regulation (EC) no. 2006/2004 and Directive 2009/22/EC. It even gave rise to a report titled «Online dispute resolution: webscraping report» available in English and French at: https://ec.europa.eu/info/online-dispute-resolution-1st-report-parliament_en

Whenever these two instruments are used to collect and organise, structuring and organising data contained in websites that can be considered databases (or disclosed therein), we are faced with their possible legal protection. In fact, here and there, we can find exclusive rights to use databases granted to their "proprietor", that limit or prohibit third-party use, including indexing and scraping.

Throughout this study, we will analyse the legal regime applicable to the use, by third parties, of databases available on websites, as legally protected assets[12]. An unparalleled European legal framework[13]. The first question to ask is: what are website databases?

Defining a database is not an easy task. Let's start with a realistic concept: a database is the name given to the structure that serves as a data repository[14].

We can give some examples of online (or website) databases, taken from European jurisprudence:

1. Listing platform for second-hand car ads, with make, model, mileage, year of manufacture, price, colour, bodywork number, type of fuel and number of doors[15];

2. Calendar of football matches with dates, times and the name of the teams related to the different matches, posted on the website[16];

---

[12] Concept of Oliveira Ascensão, José de ("Criminalidade Informática"; *in Direito da Sociedade da Informação* – Vol. II; Coimbra Editora; Coimbra, 2001; page 219) that with regard to computer goods liable to criminal offenses, lists only three: computer programs, databases and topographies of semiconductor products. The Law, however, has changed (Cybercrime Law – Law No. 109/2009, of September 15th), but the teachings remain current.

[13] On the rejection of the regime in these countries, see Vicente, Dário Moura; "A Informação Como Objeto de Direitos"*; in Revista de Direito Intelectual no. 1 - 2014*; Almedina; Coimbra, 2014; pages 126 to 128.

[14] A database must be evaluated from a transactional point of view, in the set of guarantees it offers in terms of atomicity, consistency, isolation and durability. In computer science this group of characteristics is defined by the acronym ACID (Atomicity, Consistency, Isolation and Durability). On this matter see Haerder, Theo, and Reuter, Andreas; "Principles of transaction-oriented database recovery."; *in Computing Surveys, Vol. 1, No. 4, December 1983*; Association for Computing Machinery; New York, United States of America, 1983; pages 287 to 317.

[15] As per the Decision in Case C-202/12 (Innoweb BV vs Wegener ICT Media BV and Wegener Mediaventions BV).

[16] See the Decision in Case C-444/02, of November 9th, 2004 (Fixtures Marketing vs Prognostikon Agonon Podosfairou Organisms Case).

3. Online list of poems, showing the frequency with which the poems are referred to in the different anthologies, the author, title, first line and year of publication of each poem[17];

4. List of horse races with the name, place and date of the race, with the distance to be covered, the admission criteria, the closing date of submissions, the amount of the submission fee and the maximum amount contributed by the racecourse for the prise to be delivered at the end of the race, also posted on the website[18].

Legally, the concept is broader (because it does not just refer to online databases). Legally speaking[19], a database is a collection of works, data or other independent elements, arranged in a systematic or methodical way and capable of individual access by electronic or other means. From this concept, in addition to the structure, we will only have a database if it is composed of at least one of the following three types of content:

1. Works (literary, artistic, musical or other);

2. Data (personal, market values, temperatures, traffic, *inter alia*); and

3. Independent elements (or, for us, "material"[20]).

---

[17] As per the Decision in Case C-304/07 (Directmedia Publishing GmbH vs Albert-Ludwigs-Universität Freiburg).

[18] At issue was the scraping of this database by a sports betting platform. As per the Decision in Case C-203/02, of November 9th, 2004 (The British Horseracing Board vs William Hill Organisation).

[19] See Article 1 of Directive 96/9/EC of the European Parliament and of the Council of March 11th, 1996 on the legal protection of databases.

[20] This seems to result from the wording of the diploma. In the Recitals, the legislator distinguished three terms: "works"; "data"; and "materials". Interpreting the concept of database established in Article 1 of the Directive in light of Recital (17), the "data" is part of the set of "materials", which include the concepts of: "data", "texts, sounds, images, numbers or facts". The English version is the only one to always refer to "materials", in the Portuguese, French and German versions the legislator addressed the same reality – or so it seems – with different concepts (independent and material elements). Perhaps the reason comes from a longer experience: the United Kingdom already had a database protection regime before the proposal for a directive by the European Commission in 1992. The relative confusion will arise, as it appears from the European legislative procedure, from an unease about the concept of data as material when other intellectual property agreements and conventions were being drafted. It is the European Commission itself that affirms this by adopting the new (confused) wording proposed by the European Parliament. We can read, at a certain point in the Commission's communication adopting the amended proposal for a directive that the "Parliament intended, through a series of amendments – some somewhat imprecise from a linguistic point of view…". Regarding the use of the term "data" in the concept of database, the Commission stated that "The inclusion of the word "data" in the definition of "database" represents a timely precision, consistent with the GATT's TRIPS project and with the proposal for a protocol to the Berne Convention.".

Thus, we have: works, such as intellectual creations; materials, such as texts, sounds, images, numbers, facts that are not intellectual creations or considered to be products of abstraction from the reality they materialise (meaning, non-creative reproductions or recreations such as photographs captured by surveillance systems or call recordings); and data, such as description or abstract decomposition of those materials[21-22].

To summarise it in a sentence, as regards the concept of "materials": when worked by creativity, in an original manner, they will give rise to works; when described or decomposed in an abstract manner, they will give rise to data; all others, remain unworked **materials**, yet all prone to being included in a database, as "independent elements", capable of being accessed individually and, when separated, do not affect their individual content[23].

The relevant databases for this study are online databases: collections of works, data or materials, arranged in a systematic or methodical manner and capable of individual access on publicly accessible websites.

It is also necessary to point out two preliminary notes, which we shall further analyse later on. First, not all databases (including those of the jurisprudence examples) are susceptible to protection. Second, computer programs used in the creation or operation of databases are not protected by this regime[24].

---

[21] Examples: in the text, the fonts used throughout the document; in sound, the variation in volume; in images, the resolution; in numbers, their inclusion in the base as part of a sequence; in facts, verified temperatures, the name of a particular person or the number of rooms in an apartment.

[22] Also included here is the so-called metadata, meaning, the information that defines the data. On metadata, see Pomerantz, Jeffrey; *Metadata*; The MIT Press; Massachusetts, United States of America, 2015.

[23] This is the understanding of the CJEU, for example, in the decision in Case C-444/02, of November 9th, 2004 (Fixtures Marketing vs. Prognostikon Agonon Podosfairou Organisms Case). In order for it to be considered that they are individually accessible, it is necessary, under the terms of the decision, to include a method or a system, of whatever nature, that allows each of these elements to be found.

[24] As per article 1(3) of the Directive.

## 2. Database Protection

## 2.1. Introduction

In 1996, the European legislator created a specific legal regime for the protection of original databases, a protection that still remains in force today. This was not a complete novelty – it was already relatively accepted that databases with an original and creative structure would be protected by copyright under the general rules of the Berne Convention (1886) and the TRIPS Agreement (1994), when not, in general, by Competition Law[25] –, but it was necessary (at the time and for market reasons) to legally protect non-creative databases[26]. With this in mind, an innovative[27] European regulatory framework was created, which simultaneously protects both.

This double protection[28] was introduced by Directive 96/9/EC of the European Parliament and of the Council, of March 11, 1996, on the legal protection of databases (the «Directive»), implement in Portugal by Decree- Law no. 122/2000, of July 4, amended (without relevant modification of the regime) by Law no. 92/2019, of September 4th (the «DBDL»). It provides for specific protection through copyright for creative databases, and *sui generis* rights for non-creative ones. It also provides for the protection of the legitimate user of a database, preventing contractual restrictions on its use.

---

[25] In this regard, Grosheide, F.W.; "Database Protection – The European Way"; *in Washington University Journal of Law & Policy Washington University Journal of Law & Policy - Volume 18 - Symposium on Intellectual Property, Digital Technology & Electronic Commerce*; pages 42 *et seq.*.

[26] To get a picture of the 30 months of European legislative process that preceded the approval of the regime, see Gaster, Jens L.; "The New EU Directive concerning the Legal Protection of Databases" *in Fordham International Law Journal - Vol. 20, Issue 4, January 1996; The Berkeley Electronic Press*; 1997; pages 1129 to 1150.

[27] For an analysis of the historical evolution of the regime and its study at the international level and objectives, see Hugenholtz, P. Bernt; "Something Completely Different: Europe's Sui Generis Database Right" *in The Internet and the Emerging Importance of New Forms of Intellectual Property (Information Law Series, Vol. 37)*; Wolters Kluwer; Alphen aan den Rijn, Holanda, 2016; pages 205- 222.

[28] In the words of Akester, Patrícia; *Direito de Autor em Portugal, nos PALOP, na União Europeia e nos Tratados Internacionais*; Almedina; Coimbra, 2013; page 238.

To better understand the Directive's protection system, we distinguish databases into three types[29]: "creative databases" (such as intellectual work), protected by copyright and created by their "author"; "non-creative databases", protected by *sui generis* rights and created by their "manufacturer"; and "common databases" which do not correspond to either of the previous two and which have simple contractual protection (if the creator so determines), created by their "creator".

## 2.2. Creative Databases

A creative database is necessarily an intellectual creation[30] – meaning, it has to present a special selection or arrangement of content (4(1) *ex vi* 4(2) of DBDL) designed to facilitate access and analysis information, usually dispersed, in one place or through a single platform.

According to the CJUE[31], the legislator sought to protect only those databases that are the original expression of their author's creative freedom, showcasing originality in that selection or data display - excluding from such creativity the intellectual efforts and expertise in the constitution of the database, in the meaning of the selection criteria or in the form that the data is displayed.

In short, the concept of a database protected by copyright (creative) will be the entire collection of works, data or other independent elements, arranged in a systematic or methodical way and capable of individual access by electronic or other means, which, by the selection or arrangement of the respective contents, constitute intellectual creations.

The protection granted to it is the same as for copyright and related rights (4(1) of DBDL) and shall remain in force for a period of 70 years after the death of its author or the first disclosure of the database depending on whether it is a natural person or another entity (6(1) and (2) of DBDL).

---

[29] It is not a clear distinction, but one that results from the legal regime adopted at the European level and implemented in national law. Sá de Mello, for instance, distinguishes between "database-work" and "database-product" (as per Sá de Mello, José de; *Manual de Direito de Autor e Direitos Conexos*; 4th Edition Reformulated, Updated and Expanded; Almedina; Coimbra, 2020; pages 121 to 123), also identifying them according to the legal regime applicable to each.

[30] As a specific creation of the respective author in the selection or arrangement of the materials that comprise them.

[31] See the Decision in Case C-604/10, of December 10th 2010 (Football Dataco Ltd. vs Yahoo! UK Ltd. Case).

A creative database is protected by copyright. The object of this protection is the database (as a structure), minus its content (1st part of 4(3) of DBDL).

This protection (of form) does not replace any protection (of the substance) of the content displayed therein, for example in matters of personal data protection, business secrets or copyright, which may even belong to other people (2nd part of 4(3) of DBDL).

On this structure, the Law guarantees the author the exclusive rights to[32]:

a) Permanent or temporary reproduction, by any process or form, of all or part of the database;

b) Translation, adaptation, transformation, or any other modification of the database;

c) Distribution of the original or copies of the database;

d) Carrying out any public communication, exhibition or public representation of the database;

e) Reproduction, distribution, communication, exhibition or public representation of the derived database, without prejudice to the rights of the person carrying out the transformation;

f) Mention of the name in the database and claim of authorship.

These rights are legally guaranteed by a criminal penalty, stating that the breach of these rights, namely the reproduction, disclosure or communication to the public for commercial purposes, without the author's authorisation, is punishable by imprisonment of up to 3 years or with fine (11 of DBDL).

Indexing and scraping allow, although not necessarily and as techniques for organising and obtaining data, the reproduction, translation and adaptation of online databases.

For any of those operations to be possible, the user of the creative database must correspond to a legitimate user, thus demonstrating that it is a necessary act for its use, without harming the normal operation of the database, nor causing a serious

---

[32] 7 and 8 of the DBDL.

and unjustifiable harm to the legitimate interests of the author (9 *ex vi* 10(2) of DLDB) [33-34].

These concepts, intentionally indeterminate to allow their permanent updating, have been furthered by European jurisprudence. Thus, it is now considered that the use of a *legitimate user* necessarily respects the general three-step rule[35]. This rule establishes that restrictions on copyright must be admitted, provided that the following three steps of admissibility[36] in use are followed:

a) It is exceptional: restrictions on copyright are specifically provided for by Law, so that uses under it are considered valid (restricted cases);

b) Corresponds to a common use: restrictions must not jeopardise its operation by the author, namely not compromising servers or the normal service of the websites used (normal operation of the work);

c) Respects the legitimate interests of the author of the database: restrictions must not cause serious and unjustified harm to these interests (which must be legitimate), namely by not diverting clients or changing, with financial damage, the way in which potential users interact with the database (respect for the legitimate interests of the author).

Scraping and indexing are techniques that correspond to a normal use of databases in the current context of information. When they comply with the aforementioned three-step rule, they should be considered instruments of legitimate use of copyright-protected databases.

---

[33] In this regard, on the possibility of a legitimate user perform the necessary acts to access and use the database, see Menezes Leitão, Luís; *Direito de Autor*; Almedina; 4th Edition; Coimbra, 2021; pages 349 and 351 and 352 and Akester, *op. cit. page* 241.

[34] We could not fail to state the following: article 9 of the DBDL refers to the possibility of the user to perform all acts provided by article 5 of the same decree-law that refers to the ownership of the database; this reference does not make any sense; analysing the Directive one can understand that what the national legislator did was copy the text of its article 6, which also referred to article 5 (of the Directive), this one related to the author's acts (and which corresponds to article 7 of the DBDL); interpreting this decree-law in accordance with the Directive that it implemented, a corrective interpretation is required – in article 9(1), where it says "article 5", it should be interpreted as "article 7".

[35] In this regard, see Menezes Leitão, *op. cit*. page 349. See also 10(2) and 14(2) of the DBDL.

[36] From the Berne Convention (as per 9(2)), also provided for in article 78 (4) of the Copyright and Related Rights Code, this rule also includes the protection of databases. For simplicity, see, Akester, *op. cit*. pages 118 and 119.

The use by a legitimate user is not only lawful: it is legally protected against any restriction by act or contract (9(1) *ex vi* 10(2) of DBDL), and it is not permissible for the author to prevent the use that such legitimate user makes of the database. Contract provisions that establish otherwise shall be considered null (9(2) of DBDL).

## 2.3. Non-Creative Databases

For a non-creative database to be protected, it must result from a substantial investment – in obtaining, verifying or presenting its content –, from a qualitative or quantitative point of view, on the part of its manufacturer[37].

The legislator left the concept of qualitative and quantitatively substantial investment undetermined so that it could be implemented, on a case-by-case basis, by the courts. The CJUE has already done this, although it has not yet established amounts [38]. Firstly, the investment must be relevant and made to obtain, verify or present the content of the database, not considering that which is supported for the creation of the data[39]. Secondly, this investment can be quantitatively or qualitatively relevant. Quantitative appraisal is carried out through calculation (financial investment), while qualitative appraisal is carried out in relation to non-quantifiable efforts (intellectual effort; energy expenditure) [40].

---

[37] 12(1) DBDL.

[38] See The British Horseracing Board Case *op. cit*.

[39] Although this is a brief study, we cannot fail to draw attention to this point. The CJUE (*ibid.*) considered that the investment with "creation" of data (but only with acquisition, verification and presentation) is not accounted for. This is usually seen in data generated by a main activity which is then entered into a database as a secondary or necessary activity. This is the case of the so-called "spin-off" databases, generated as a necessary by-product of the main activity of their manufacturer. According to the spin-off database theory, in its broad version, no spin-off database would be protected. This theoretical formulation was apparently rejected by the CJUE (thus Lohsse, Sebastian; Schulze, Reiner; e Staudenmayer, Dirk; "Trading Data in the Digital Economy: Legal Concepts and Tools"; *in Münster Colloquia on EU Law and the Digital Economy III*; 1st Edition; Nomos Verlagsgesellschaft; Baden-Baden, Alemanha, 2017; page 29, note 10), as even a spin-off database can still be protected (according to the decision) if there is a substantial investment in its presentation. Or so it seems. Against this position, though not substantiating it, see Davison, Mark; "Database Protection: Lessons from Europe, Congress, and WIPO" *in Case Western Reserve Law Review Vol. 57*; page 838. The subject of spin-off databases theory is very interesting and plays a very important role in the good regulation and prevention of information monopolies, but it is also too extensive to be addressed here.

[40] This is the interpretation of the CJUE of Recitals 7, 39 and 40 of the Directive, see decisions in Cases No. C-46/02 (Fixtures Marketing Ltd v. Oy Veikkaus Ab), C-338/02 (Fixtures

A non-creative database that results from a qualitatively or quantitatively relevant investment is protected by *sui generis* rights. The object of this protection is the content of the database[41], as an exclusive right of use, against third party uses that cover all or quantitatively or qualitatively substantial part of this content (the user is only allowed to use a non-substantial part).

The Law protects this content by reserving the manufacturer the right to authorise or prohibit[42] any act of appropriation, in whole or in substantial part, of the content of the database (extraction[43]) and against any act of public use or disclosure, of all or of a substantial part, of the content of a database (reuse[44])[45].

As with creative databases[46], the *legitimate user* of a non-creative database can also perform all acts, namely extracting and reusing data from that database, provided that this is done in a way that does not harm the normal operation of the database, nor cause serious and unjustifiable harm to the legitimate interests of the author or holders of copyright and related rights on works and services incorporated therein, and as long as these operations only address the non-substantial part of the database (14(1) and (2) of DBDL) and in a non-repeated manner[47] (12(6) of DBDL).

---

Marketing Ltd v. Svenska Spel AB), and No. C-444/02 (Fixtures Marketing Ltd v. Prognostikon Agonon Podosfairou Organisms AE (OPAP)), both from November 9th, 2004.

[41] The investment made has been defended as the object of protection (in this regard, see Menezes Leitão, *op. cit.* page 350), but most of the doctrine has not directly addressed the matter. This is also the position of the European Commission, contained in its assessment reports on the regime, the first in 2005 ("DG INTERNAL MARKET AND SERVICES WORKING PAPER – First evaluation of Directive 96/9/EC on the legal protection of databases") and the second in 2018 (COMMISSION STAFF WORKING DOCUMENT – Evaluation of Directive 96/9/EC on the legal protection of databases; SWD (2018) 147 final).

[42] The Directive is only about authorising. The DBDL's use of the word "prohibit" is very unfortunate. A ban would only be justified in two situations: a) when the Law allows the user to use the database; b) when the Law does not allow it, and there must be an authorisation. It turns out that in the first situation, it is the Law itself that does not admit any contractual restriction (the aforementioned prohibition), and in the second it would not make sense to speak of prohibition, but removal of authorisation.

[43] 12(2) of DBDL: "extraction" means the permanent or temporary transfer of all, or a substantial part, of the content of a database to another medium, by any means or in whatever form.

[44] 12(2) DBDL: "reuse" means any form of distribution to the public of all, or a substantial part, of the content of the database, namely through the distribution of copies, rental, online transmission or any other form.

[45] 12(1) DBDL.

[46] As per 2.2. above.

[47] According to the CJEU, this prohibition serves to avoid circumventing the prohibition on the use of a substantial part (as per the Decision in The British Horseracing Board Case *op. cit.*). In other words, it only takes place when the non-substantial use, due to its systematic nature, results (in the end) in a substantial use.

The use by a *legitimate user* is legally protected against any restriction by act or contract (14(1) of DBDL), and the author cannot prevent the use of the database by such a *legitimate user*. Any contractual provision that established otherwise shall be considered null (14(3) of DBDL).

The protection is maintained for a period of 15 years, starting from the 1[st] day of January of the year following the public disclosure of the database (16t(1) of DLDB). In any case, the disclosure must be made within a period of 15 years from the 1[st] day of January of the year following the date of the creation of the database, under penalty of it expiring (16(2) of DBDL).

This 15-year period is renewed whenever the database is substantially modified - quantitatively or qualitatively - (including that resulting from the accumulation of successive additions, deletions or amendments), which leads to a new initial investment (17 of DBDL).

## 2.3.1. Also: the importance of links

Hyperlinks (links) have, *per se*, a neutral legal relevance. They are generally allowed[48] being considered as a simple access[49] to a certain website (and to the work or database made available therein), thus falling within the disclosure sought by the website's owner when placing the contents online.

On the other hand, the use of hyperlinks (or links), when reproducing part of a non-creative database, is relevant, as it can prevent or cause damage to the database creator. Contributing to fulfilling or breaching the three-step rule.

From the outset, the use of links to the origin website can avoid the damage eventually caused by the loss of customer traffic, when it directs users from another website back to the original website (simple links). On the contrary, when they reproduce the entire content of the database to which they refer (for example: a link associated with all relevant information by image), there is a risk of causing the user to lose interest in visiting that original source. Furthermore, sending the user to sub-pages of the original website may result in loss of revenue associated with the entry of the user via the homepage (for example: ads or campaigns only present on the homepage).

---

[48] In that regard, Menezes Leitão, *op. cit.* page 366 and 367. This way of disclosing access to information shall only be considered unlawful insofar as it serves an unlawful purpose (for example: intentionally directing the user to a page with illegal contents for the purpose of distributing them).

[49] In the words of Oliveira Ascensão, who argues that it is merely an "access to someone else's site" (see *Propriedade Intelectual e Internet*, pages 14 and 15; available at http://www.fd.ulisboa.pt/wp-content/uploads/2014/12/Ascensao-Jose-PROPRIEDADE-INTELECTUAL-E-INTERNET.pdf).

The use of simple links will therefore be preferable in principle.[50].

## 2.4. Common Databases

Ordinary databases are, in short, non-creative databases that did not require a substantial investment in obtaining, verifying or presenting their content. We are, therefore, before databases that the legislator chose not to protect – not attributing a special creative (copyright) or economic (*sui generis* rights) value to it[51].
Thus, the only protection that can be found for these databases is contractual. But this is also limited.

First, there are situations in which it is not permissible to place restrictions on the use of the database[52]. These are the cases in which the database is considered to be protected (by copyright or *sui generis* rights) and the use made of such database complies with the three-step rule[53] – the *legitimate user*. All restrictions and contractual penalties shall be generally considered null and void - as per article 294 of the Portuguese Civil Code ("CC") *ex vi* of article 9(2) and article 14(3), both of the DBDL –, having no effects.

---

[50] In this direction, the decision of the CJUE in the Innoweb Case *op. cit.* By "simple links" we mean the distinction between deep links (directing the user to the website's internal pages) and simple (containing the link to the relevant website's home page).

[51] We will not mention here the Freedom of Information Principle, the violation of which could call into question the very validity of a contractual restriction. In fact, it is this same principle that determined the opening of limitations to copyright and *sui generis* in relation to databases. For a better understanding of this principle and the implications it has for the regulation of databases, such as sets of information, see Oliveira Ascensão, José; "*Bases de dados electrónicas: o estado da questão em Portugal e na Europa*"; *in Direito da Sociedade da Informação - Volume III*; Coimbra Editora; Coimbra, 2002; pages 17 to 18 and 21 to 26).

[52] As per 2.2. and 2.3. above.

[53] See 2.2. above.

In the case of common databases, we see find it very difficult to argue for the legal relevance of their protection – meaning, contractually. First, because the legislator decided not to protect them. Furthermore, if the requirements of the three-step rule are met (being sufficient the lack of serious and unjustifiable harm to the interest of its creator, and the maintenance of the normal operation of the website), the extensive interpretation of the Law must be admitted (*a maiori, ad minus* argument), to conclude that if the Law protects the *legitimate user* against restrictions of the author or manufacturer of a legally protected database (allowing the most), it shall also necessarily do so for databases that do not deserve legal protection (will allow the least). We must conclude the following: contractual restrictions on the use of a publicly accessible common database should not be permissible when we are dealing with a legitimate use that does not harm the legitimate interests of its creator and that does not compromise its normal operation[54].

Outside of this context of legitimate use or when the common database is not publicly accessible, the contractual restriction of use by third parties or counterparties, namely indexing or scraping, should be considered fully acceptable, fully applying the general private autonomy principle[55-56], being the parties free to regulate their relationship when using the database.

---

[54] See the apparently conflicting decision of the CJEU in Case C-30/14 (Ryanair Ltd vs PR Aviation BV).

[55] Regarding this principle see Carvalho, Jorge Morais; *Os Limites à Liberdade Contratual*; Almedina; Coimbra, 2016; pages 10 *et seq.*

[56] Also the CJUE in the Ryanair Case *op. cit.*

## 3. Crowdsourced Databases[57]

At this point we propose a new qualification of databases - the "crowdsourced databases". As we will see below, we propose a category of databases that is not expressly provided for by law, but which results from its interpretation and application to a set of identical (or typical) concrete cases: a *type* of database that cannot be subject to *sui generis* rights.

Crowdsourced databases are (as the name suggests) databases filled by the collective contribution of third parties, on a website freely accessible to the public. Because it differs from the others solely because of its content, this type of database is relevant only in terms of *sui generis* rights[58]. On the positive side, it is a type of database that gives the user ample autonomy. On the negative side, it is a legal concept that restricts the scope of *sui generis* rights, hence excluding all those that fit in this type.

Crowdsourced databases present a set of common elements that suggest this typological qualification: i) the prevalence of the collaborative framework in obtaining and verifying data; ii) the interest of a third party in contributing, aiming at divulging the content; iii) the existence of an economic interest of the underlying creator, independent of the use of the database.

The first and most differentiating factor is the collaborative regime in obtaining data, not being the owner of the database who obtains the data (or the majority of the data), but someone else. In other words, this effort is handed over to third parties, whether these are indeterminate (public) or a certain group. The contractual relationship between the owner and the third party can take the most diverse forms: the owner may be providing an advertisement service (the third party contributing with the advertisement content) or offering access to the database as an advantage to a franchisee or agent, to which such franchisee or agent also contribute.

---

[57] At the time of writing this study, we are not aware of any other study that addresses the existence of this type of database or that creates a similar category.

[58] Since it only the content is relevant, the qualification to be given to its structure and, consequently, its possible protection by copyright is irrelevant. A crowdsourced database may or may not be protected by copyright.

In any of these cases, we will always find an interest for the third party to contribute. There will be many cases in which the third party pays to make this contribution – this is the case of advertising platforms for auctions and proposals to the public for the purchase and sale or lease of goods. As these are publicly accessible databases, the contribution of the third party is functionally linked to the objective of disclosing the contributed content. The purpose pursued by the parties, owner and third contributing party, is the disclosure of said content to the public.

In these databases, the economic interest of the creator is not related to the access and use of the data contained in the database, but to the offer of this space to third contributing parties. The investment is made only for this purpose, looking for a direct return - when it sells the possibility to contribute (example: advertising platform for the sale of goods) - or indirect - when it offers this possibility as a benefit in another contract (example: advantage for franchisees with use of a platform to display the goods for sale).

We may also find, although it is not necessary for its qualification, the application of the collaborative method in the verification of data - this is always the case when the third party that contributes to the obtainment of data also takes on the task of verifying the obtained data.

The combination of these three traits is crucial in qualifying the relevance of the investment for non-protection purposes[59]. As we saw above (2.3.), for a non-creative database to be protected, it must result from a substantial investment in obtaining, verifying or presenting its content.

It is precisely with the evaluation of these three investment objects (obtainment; verification; presentation) that we better understand each of those three traits (collaborative regime; interest of the third party; interest of the creator).

First, there is typically no substantial investment in crowdsourcing data. The burden for obtaining data is handed over to third parties. When the owner of a database bears the cost of obtaining data from the third party, the owner does not operate a database in a crowdsourced regime. These will be the cases in which the third party receives a(n) (effective) remuneration for submitting contributions.

---

[59] 2.3. above.

Secondly, we shall also not find a relevant investment in verifying the data entered. It may happen that the creator takes on a residual verification role, which does not satisfy the requirement of substantiality from a qualitative point of view. Thus, it is up to the third party to ensure that the contribution made is adequate or, in other cases, it is up to the group of third parties to carry out such assessment. It may even be the case that there is no verification, and the database users are responsible for the validation effort - this is the case for sales platforms based on comments or evaluations from users and customers.

Finally, thirdly, there is also no substantial investment in data presentation. In fact, this investment is made to attract third parties who contribute, having the database as a structure, and not to present the database, as a result with already filled in content - now, the Directive intends to protect those owners who, even if they do not invest in obtaining and verifying data, invest in its presentation in order to better disclose the data, not, as is the case here, in order to capture more contributions from third parties.

In conclusion, for the consequences. These databases should be seen as channels for publishing third-party contributions. Which means that their non-protection does not harm the creators. Firstly, because it does not directly affect their economic interest (based on the contribution and not on the use made of it). But there is also no damage, even if indirect, since third parties are not looking for an exclusive channel, but only a channel they consider to be relevant. The use made of this database, even if by reproduction or adaptation, is a way of expanding the distribution channel. In conclusion, even without protection by the Directive, the database is created and maintained by the owner and filled in by contributions from third parties, which does not justify a reinforced protection such as that resulting from *sui generis* rights. However, the misuse that damages the normal operation of the database or affects the return on investment of its creator is still deemed illegal[60].

---

[60] It would represent, at the very least, an abuse of rights.

Finally, considering that crowdsourced databases are, for protection purposes, common databases, we refer to our previous position regarding the protection of the legitimate user of this type of databases (see extensive interpretation exercise – 2.4. above): contractual restrictions on its use are not acceptable when it is publicly accessible, and the use made by the user does not harm the legitimate interests of its creator or compromise its normal operation.

From an economic perspective (of consequences), we must point out the irrelevance of the issue, knowing that it has not been possible to show that the opposite hypothesis (maintaining the current protection regime) will have a more positive economic impact[61]. Our opinion is based solely on legal and technological arguments: the opening of these databases, in addition to a social and ethical requirement, brings liquidity to the information market, not creating any disincentive (as the return of manufacturers is not affected) to the creation of more databases. The consequences can only be positive, in all dimensions.

We recall that the European legislator thought it necessary to protect the return on investment - when considerable[62] - of the manufacturer, against possible misappropriation of the result of its effort and the risks it undertook[63], defending it from the abusive use of "parasitism" [64] It chose to do so, without ignoring that information has an increasingly relevant role in society, being its management increasingly relevant, considering its value and its exponential growth in size[65] and that this regime did not prevent the application of competition law, namely the protection of the manufacturer not being able to prevent the creation of new databases with the same information or encourage the formation of information monopolies[66].

---

[61] See the European Commission's 2005 and 2018 evaluation reports above. In particular, and as a result of the 2018 conclusions: "there is no evidence to conclude that the sui generis right has been fully effective in stimulating investment in the European database industry, nor in creating a fully functioning access regime for stakeholders" (page 46).
[62] Recital 7.
[63] Recitals 39 to 41.
[64] Recital 42.
[65] Recitals 9, 10 and 11.
[66] Recital 47.

At the time, the example of a digital database came in the form of a CD[67]. Thus, it is no surprise that we have the need to create different categories from those that result directly from the text of the Directive, as results from our Introduction. On the other hand, the legislator knew that reality would evolve in the unpredictable similarity of what had been the technological evolution until then, having, for this reason, resorted to indeterminate concepts that constitute flexible constructions of protection principles, which can be perfectly applied to this new category of databases, keeping the text of the Directive up to date.

In order to anticipate what would be the fair case decision in several similar cases, we tried to typify this similarity and we now propose this new legal category of databases: crowdsourced databases.

---

[67] See Commission Report (2018; *op. cit.*; page 37): "The proposal for a Database Directive was first adopted in 1992, with the CD-ROM market as reference".